\documentclass[12pt]{article}

\usepackage[utf8]{inputenc} 
\usepackage{amsmath, amssymb} 
\usepackage{graphicx} 
\usepackage{hyperref} 
\usepackage{geometry} 
\usepackage{float}
\usepackage{authblk}

\title{Charging a dimerized quantum XY chain}

\author[1,2]{Riccardo Grazi}
\author[1,2]{Fabio Cavaliere}
\author[1,2]{Niccolò Traverso Ziani}
\author[1,2]{Dario Ferraro}

\affil[1]{\small Dipartimento di Fisica, Università degli Studi di Genova, via Dodecaneso 33, 16146, Genova, Italy}
\affil[2]{\small CNR SPIN, via Dodecaneso 33, 16146, Genova, Italy}
\date{}
\begin{document}

\maketitle

\begin{abstract}
Quantum batteries are quantum systems designed to store energy and release it on-demand.
The optimization of their performance is an intensively studied topic within the realm of quantum technologies. Such optimization forces the question: how do quantum many-body systems work as quantum batteries? To address this issue, we rely on symmetry and symmetry breaking via quantum phase transitions. Specifically, we analyze a dimerized quantum XY chain in a transverse field as a prototype of energy storage device. Such model, which is characterized by ground states with different symmetries depending on the Hamiltonian parameters, can be mapped onto a spinless fermionic chain with superconducting correlations, displaying a rich quantum phase diagram. We show that the stored energy strongly depends on the quantum phase diagram of the model, when large charging times are considered.
\end{abstract}

\section{Introduction}
Models of interacting spins have great importance in various contexts of physics~\cite{De16}. Firstly, they enable quantitative predictions within the theory of magnetism~\cite{Auerbach12, Wysin15, cialone2017tailoring, cialone2020comparative}. Materials well described by spin models in various dimensions are indeed not rare~\cite{Manousakis91, Bramwell98, Mydosh_2015, Pan17, Cavaliere2023, Eckhardt2022}. At the same time, they are at the core of topological phases, both for the thermal and the ground state properties~\cite{XiaoYong07, Duivenvoorden13, Dubinkin19}. In this context, their relation to the fascinating field of frustrated systems is worth to be mentioned~\cite{Balents10, Diep13, Sacco24, maric2020quantum, lacroix2011introduction}. Moreover, they have recently played a crucial role in the determination of novel paradigms in the theory of thermalization and information spreading through the general procedure called quantum quench~\cite{Mitra18, Essler16, Porta18, Porta20}. A quantum quench is a sudden variation of a parameter of the Hamiltonian of a quantum system. Quantum quenches can be achieved experimentally within very diverse platforms~\cite{Faure19, faure2018topological, kinoshita2006quantum}.\\
Together with the large range of physical applications, many effective techniques have been put forward to analyze spin systems. These range from the celebrated spin wave theory based on the Holstein-Primakoff transformation~\cite{Holstein40}, to numerical ones, even based on the Fedotov-Popov transformation~\cite{popov1988functional, Traverso23clock}. Additionally, one dimensional models can be analyzed by means of bosonization~\cite{Ziani17, rodriguez2020relaxation, Gambetta15, TraversoZiani_2013, cryst11010020, Bloch08, Cazalilla11, trotzky2012probing, Imambekov09, Imambekov12, Wu06, Fiete06, Fiete07, Matveev07, JVoit_1995, Haldane81, Haldane_1981, giamarchibook}, as long as the low energy properties are concerned, and, even more remarkably, they sometimes might be exactly solved by means of the Bethe Ansatz technique~\cite{bethe1931theorie} or more simply by the Wigner-Jordan transformation~\cite{Jordan28}. Such exact solutions, which are at the heart of the so called theory of quantum integrable systems~\cite{Franchini17}, can indeed help in understanding some general principles of quantum many-body interacting systems and are hence extremely valuable.\\
Indeed, the existence of such techniques has strongly favored the adoption of quantum spin chains in the context of quantum technologies~\cite{Bayat22}. Indeed, a two-level system can mathematically be described as a spin 1/2, so that one-dimensional collections of qubits are immediately described by interacting spin chains. This fact has been exploited in the fields of quantum information, quantum communication, quantum computation, and more relevantly for the present article, quantum energy storage.\\
In this context, quantum spin chains have been proposed as quantum batteries (QBs)~\cite{Le_2018, Liu21, Zhao21, Catalano23, Grazi24, Ali24SuperExt}. Conceived for the first time a decade ago~\cite{Alicki_2013}, QBs are miniaturized device able to store and deliver energy in an efficient way exploiting purely non-classical effects such as quantum superposition and entanglement~\cite{Bhattacharjee21, Campaioli23, Quach23}. The interest in such devices is not only motivated by fundamental issues in quantum thermodynamics~\cite{Benenti17, 
Esposito09, Campisi16, Vinjanampathy16, Potts24}, but is also driven by the need to integrate quantum energy providers to improve the performance of quantum circuits~\cite{Hu_2022, Gemme23, Dou23, Razzoli24,Cavaliere24} or quantum computing architectures~\cite{Chiribella21, Menta24,  Elyasi2024} and by the hope that QBs could outperform their classical counterparts~\cite{Alicki_2013, Binder_2015}.\\
Among all the mechanisms that can be envisioned to charge a QB, one is tightly connected to the vast literature on quantum quenches. Such protocol works in the following way~\cite{Andolina18, Crescente22}: a system is prepared in the ground state of a Hamiltonian $H_0$, called the battery Hamiltonian. Then, the Hamiltonian suddenly changes, say at $t=0$, to a different operator $H_1$-the charging Hamiltonian- that implements the time evolution. Finally, after a time interval $\tau$, the Hamiltonian is brought back to the original operator $H_0$. In such a way, for fixed $H_0$ and $H_1$, one can evaluate the energy transferred to the quantum system as a function of the charging time $\tau$. A QB working according to the mechanism just described and composed by a collection of interacting two-level systems is usually called a spin QB.\\
Recently, it has been shown that a dimerized XY spin chain used as a spin QB shows a rich phenomenology~\cite{Grazi24}. The energy stored as a function of $\tau$ is characterized by the presence of three regimes: (i) A short charging time regime, where prominent oscillations as a function of $\tau$ are present; (ii) A plateau regime- resembling the prethermalization plateau of the quantum quench theory- where no oscillations as a function of $\tau$ are present; (iii) a recurrence time where oscillations are restored for times large enough to probe the level spacing. This last regime is pushed to $\tau\rightarrow\infty$ in the thermodynamic limit. Among the three regimes, the second one shows an interesting behavior: the energy stored strongly depends on the quantum phase diagram of the charging Hamiltonian, and in particular it is sensitive to a quantum phase transition (QPT) induced in the model by the dimerization strength~\cite{Grazi24}. In other words, in a specific case, it has been shown that the dependence of the stored energy on the parameters of the Hamiltonian when the battery Hamiltonian $H_0$ and the charging Hamiltonian $H_1$ are kept within the same quantum phase are strikingly different with respect to the case in which $H_0$ and $H_1$ are in different quantum phases. In a sense, the QPTs of the charging Hamiltonian appear to be manifest in the behavior of the stored energy.\\
In this article, we extend the analysis to a richer model, that is the dimerized XY chain in an external transverse field~\cite{PERK1975319}, to prove that, even within this larger model, the energy stored in the plateau regime strongly depends on the quantum phase diagram of the charging Hamiltonian. 

The rest of the article is structured as follows. In Sec.2., we introduce the model, its solution, and its phase diagram. In Sec.3. we introduce the charging energy and discuss its behaviour crossing various QPTs. In Sec.4., finally, we draw our conclusions. 

\section{Model}\label{Section model and diagonalization}
The model under investigation is the so called dimerized XY quantum chain in a transverse field~\cite{PERK1975319}. The Hamiltonian $H_B$ is given by
\begin{eqnarray}
     H_B &=& -J \sum_{j=1}^N \left[1 - (-1)^j \delta\right]\times
     \left[\left(\frac{1+\gamma}{2}\right) \sigma_j^x \sigma_{j+1}^x+\left(\frac{1-\gamma}{2}\right) \sigma_j^y \sigma_{j+1}^y\right] 
     +hJ \sum_{j=1}^N \sigma_j^z, 
     \label{Tesi_Ham}
\end{eqnarray}
where the subscript $B$ indicates the fact that this system will be considered as a QB. 

In the above expression, $N$ is the number of unit cells, considered here to be even for simplicity, $\sigma_j^\alpha$ (with $\alpha = x,y$) are Pauli matrices, in the usual representation, corresponding to the $j$-th site spin. The parameter $J$ is the energy scale of the system, $\gamma$, $\delta$, and $h$ characterize the strength of the anisotropy, the dimerization, and the external field respectively.

\begin{figure}[H]
    \centering
    \begin{minipage}{0.49\textwidth}
        \centering
        \includegraphics[width=\textwidth]{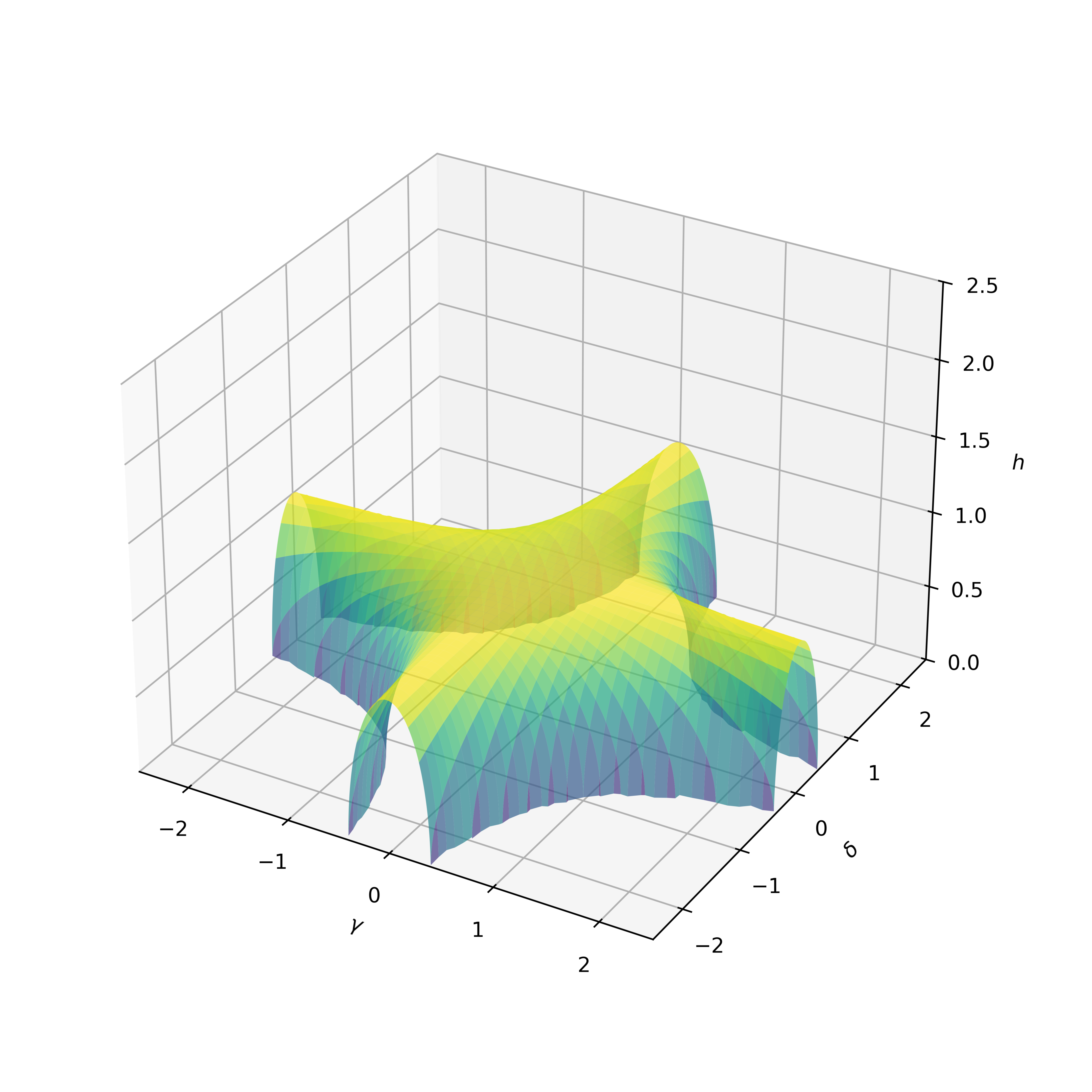}
    \end{minipage}
    \hfill
    \begin{minipage}{0.49\textwidth}
        \centering
        \includegraphics[width=\textwidth]{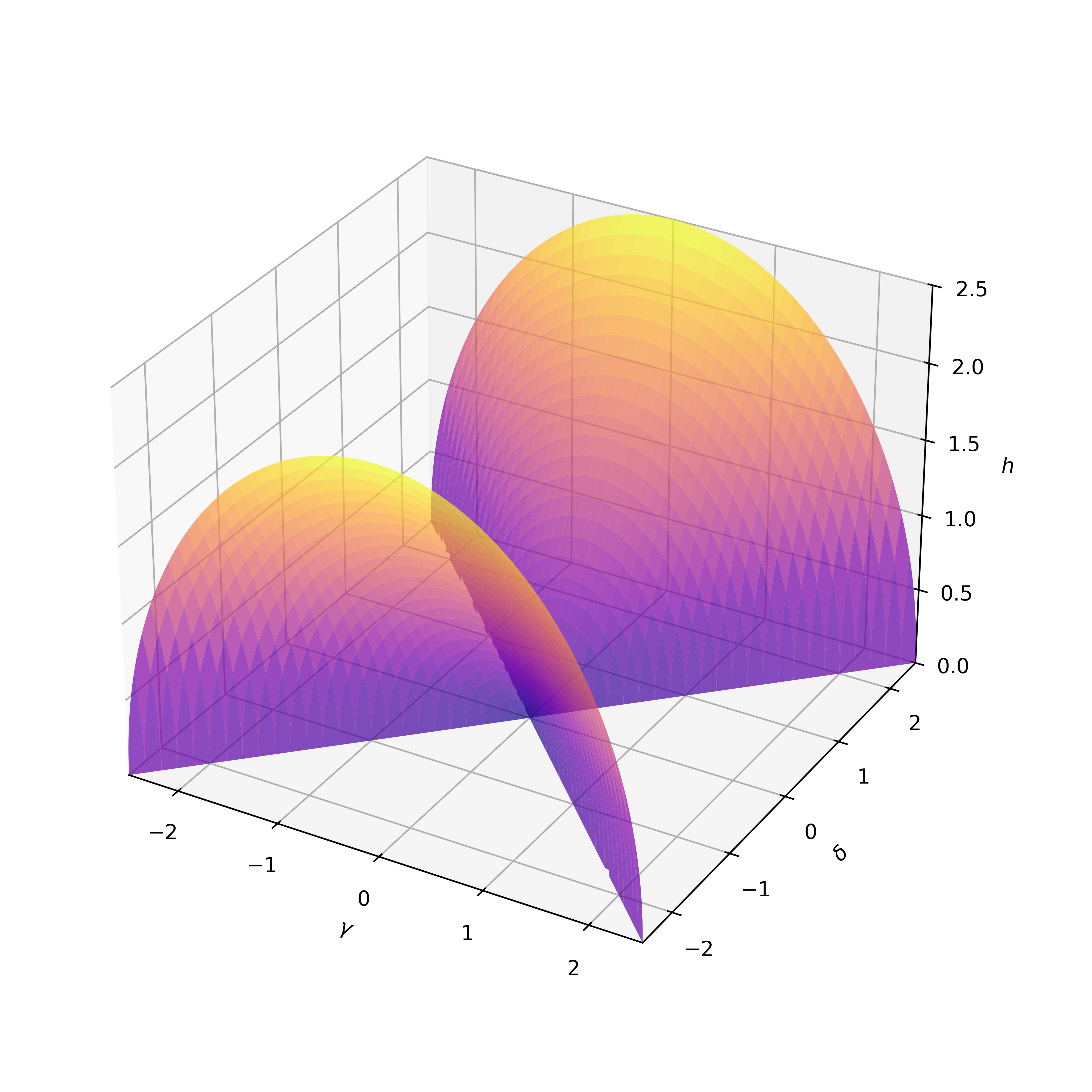}
    \end{minipage}
    \caption{3D plot of the phase boundaries $h^2 = 1 - \gamma^2 \delta^2$ (left panel) and $h^2 = \delta^2 - \gamma^2$ (right panel).}
    \label{Ph_Bound_Plots}
\end{figure}

\begin{figure}[H]
    \centering
    \includegraphics[width=0.6\linewidth]{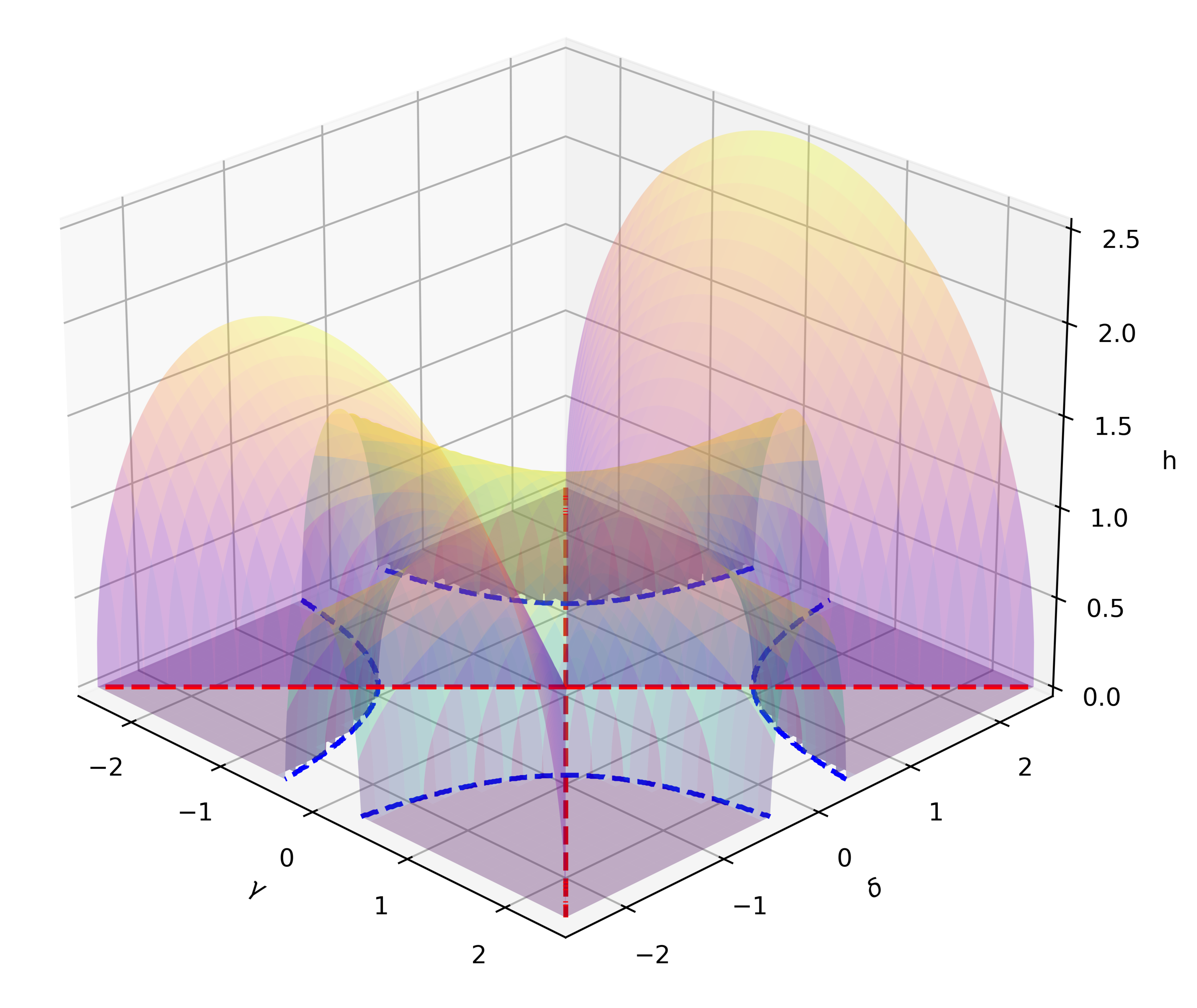}
    \caption{3D plot of the phase diagram of the model. The dashed blue and red lines are the intersections between the boundaries given by Eq.\eqref{Hyp_Line} and Eq.\eqref{Cone_Line} and the $h=0$ plane.}
    \label{h_0_QPT}
\end{figure}

By focusing for simplicity on the even-parity sector of the model~\cite{Porta20}, and considering the periodic boundary conditions $\sigma_{j+N}^\alpha \equiv \sigma_j^\alpha$, it is possible to diagonalize the above Hamiltonian by means of a standard Wigner-Jordan transformation mapping spins into free spinless fermions \cite{Jordan28}. Notice that that the obtained fermionic Hamiltonian corresponds to the dimerized Kitaev chain and is characterized by interesting topological features~\cite{wakatsuki2014fermion}, due to the competition between fractional solitons \cite{Ziani20, Jackiw76, Kivelson82, Goldstone81, Heeger88, qi2008fractional} and Majorana fermions \cite{Fleckenstein21, traverso2024emerging, Traverso22, Traverso22role, AYuKitaev_2001, Deng_2016, Prada_2018, Fleckenstein_2018, Dibyendu_2013, Flensberg_2010, Law_2009, Prada_2020, Ivanov_2001, Marra_2022, Leijnse_2012, Nayak_2008, Kitaev_2003, Oreg_2010, Lutchyn_2010}. In terms of two species of auxiliary fermionic annihilation operators $a_q$ and $b_q$, and assuming from now on $J=1$ as reference scale for the energies, we obtain (from now one we will consider $\hbar=1$)
\begin{equation}
    H_B = \sum_{q \in \Gamma} \left[\omega_{1,q} \left(a_q^\dag a_q - \frac{1}{2}\right) + \omega_{2,q} \left(b_q^\dag b_q - \frac{1}{2}\right)\right]. \\
\end{equation}
Here, $\Gamma = \{\frac{1}{2},\frac{3}{2},\frac{5}{2}, ... , \mathcal{N} - \frac{1}{2} \}$ and
\begin{equation}
    \omega_{1/2,q} = \sqrt{4h^2 + |\mathcal{Z}|^2 + |\mathcal{W}|^2 \pm 2\sqrt{4h^2|\mathcal{Z}|^2 + 16 \gamma^2 \delta^2}},
\end{equation}
where the following definitions have been used:
\begin{equation}
    \begin{aligned}
        &\mathcal{Z} = -\left[(1+\delta) + (1-\delta) e^{-ik}\right] \\
        &\mathcal{W} = -\gamma\left[(1+\delta) - (1-\delta) e^{-ik}\right]
    \end{aligned}
\end{equation}
and $k = \frac{2\pi}{\mathcal{N}}q$, with $\mathcal{N} = N/2$ the number of dimers. For $k = 0$, the gap closes at
\begin{equation}
    h^2 = 1 - \gamma^2 \delta^2, \label{Hyp_Line}
\end{equation}
while for $k = \pm \pi$, the gap closes at
\begin{equation}
    h^2 = \delta^2 - \gamma^2. \label{Cone_Line}
\end{equation}
These equations represent the phase boundaries of the model, whose plots are reported in Fig.\ref{Ph_Bound_Plots}. In \cite{Grazi24} it has been shown that the charging process at zero external field shows signatures of the presence of the QPT lines of the model. Those lines can be obtained by setting $h = 0$ in Eq.\eqref{Hyp_Line} and Eq.\eqref{Cone_Line}, as shown by the blue and red dashed lines in Fig. \ref{h_0_QPT} respectively. The Figure also show a very interesting and intricated phase diagram at finite $h$.

The goal of this work is to explore the charging process of a QB described by the Hamiltonian in Eq.~(\ref{Tesi_Ham}): the presence of a non-zero field enriches the previously studied quantum phase diagram, opening up the possibility of investigating different QPT scenarios.

\section{Study of QPT effects}\label{QPT}
The general approach of our analysis is as follows: we first select a plane in the ($\gamma$, $\delta$, $h$)-space by imposing a constraint on the model parameters. The QPT lines are then identified as the intersections of the phase boundaries, defined by Eq.\eqref{Hyp_Line} and Eq.\eqref{Cone_Line}, with the chosen plane. Next, we perform a sudden quench of one or more model parameters, focusing on the asymptotic limit where the charging time $\tau$ approaches infinity, and finally, we plot the energy stored per dimer in the battery as a function of the pre-quench parameter. In the asymptotic regime we are studying, the energy stored in the QB assumes the form~\cite{Grazi24}
\begin{equation}
    \Delta E(\gamma, \delta, h) = \sum_{q \in \Gamma} \{2\omega_{1,q} ~|\mathcal{M}_{3,1}|^2 |\mathcal{M}_{3,3}|^2 + 2\omega_{2,q} ~|\mathcal{M}_{4,2}|^2 |\mathcal{M}_{4,4}|^2 \} \label{DeltaE}
\end{equation}
Here, $\mathcal{M}_{i,j}$ represents the $(i,j)$-th element of the matrix $\mathcal{M}$, defined as $\mathcal{M} = \mathcal{V}^{-1} \mathcal{U}$. In particular, $\mathcal{U}$ and $\mathcal{V}$ are both 4x4 matrices: the columns of $\mathcal{U}$ correspond to the eigenvectors of the QB Hamiltonian, which depends on the pre-quench parameters, and similarly, the columns of $\mathcal{V}$ correspond to the eigenvectors of the charging Hamiltonian, which depends on the post-quench parameters. Equation \eqref{DeltaE} provides the value of the stored energy given a generic sudden quench of the parameters. As mentioned, our aim here is to explore the effects of QPTs. In the following, we will provide paradigmatic examples in this direction which well represent the whole phase space.

\subsection{Quench of $\gamma$ at given $h$ and $\delta$}
\begin{figure}[H]
    \centering
    \includegraphics[width=0.5\linewidth]{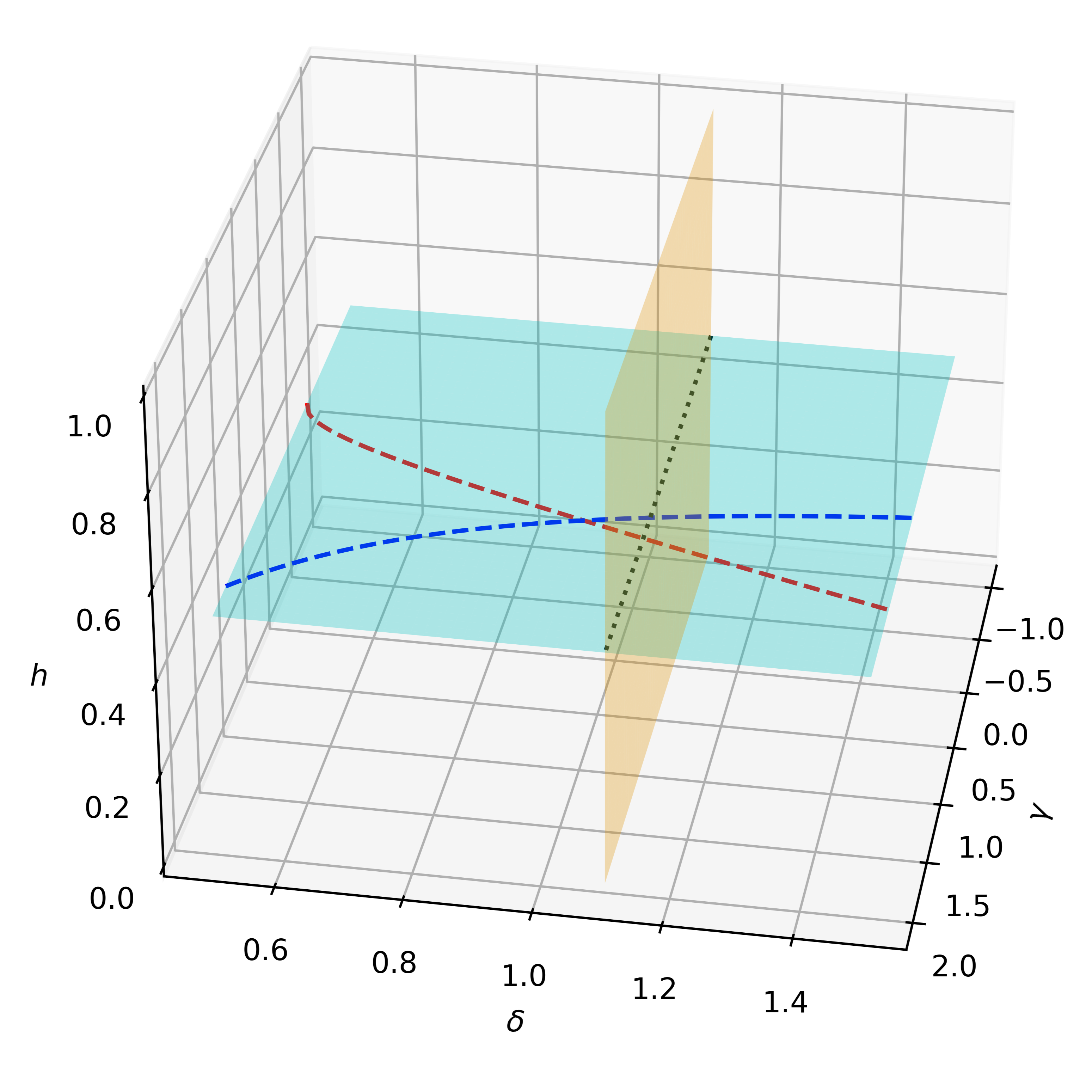}
    \caption{Section of the phase diagram of the model. The cyan and orange planes have equations $h = 0.5$ and $\delta = 1.1$ respectively, while the black dotted line is given by the intersection of the aforementioned planes and it is the line where the quench takes place. The blue and red dashed lines are the QPT lines given by the intersection of the boundaries reported in Eq.\eqref{Hyp_Line} and Eq.\eqref{Cone_Line} with the $h = 0.5$ plane.}
    \label{gamma_quench}
\end{figure}

\begin{figure}[H]
    \centering
    \includegraphics[width=0.6\linewidth]{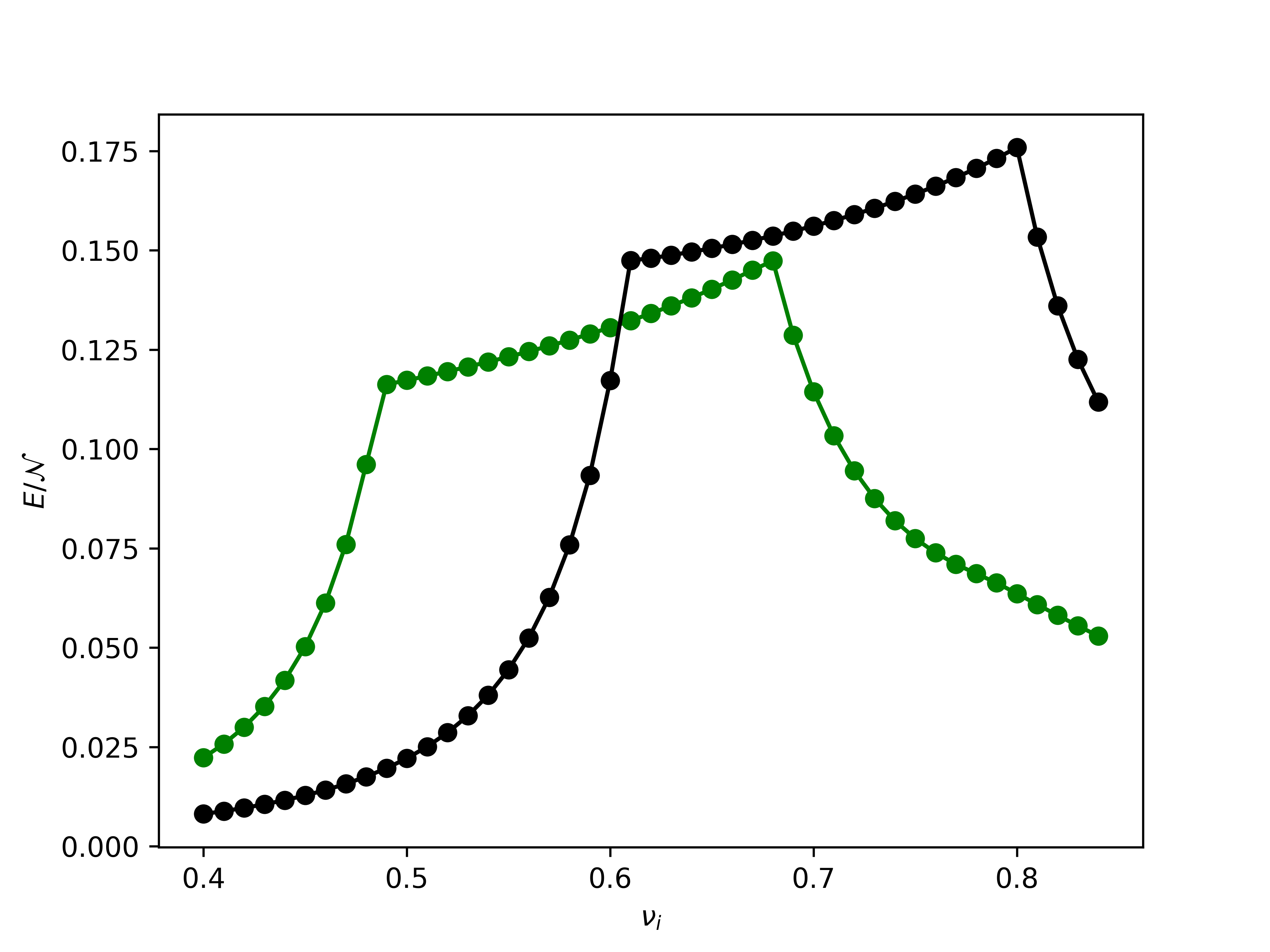}
    \caption{Energy stored per dimer as a function of $\nu_i$ after a quench of the anisotropy parameter at fixed $\nu_f = 0.3$ and $\delta = 1.1$ in the asymptotic regime, i.e. $\tau \to \infty$, 
    for $h = 0.5$ (green curve) and $h = 0$ (black curve)}
    \label{gamma_quench_RESULT}
\end{figure}

In this scenario, we perform a quench of the anisotropy parameter from an initial value $\nu_i$ to a final value $\nu_i + \nu_f$, with $\nu_i$ and $\nu_f$ positive constants, along the line given by the intersection of the $h = 0.5$ plane and the $\delta = 1.1$ one. The situation is sketched in Fig. \ref{gamma_quench}.
Figure \ref{gamma_quench_RESULT} shows the energy stored per dimer (in units of $J=1$) as a function of $\nu_i$, with $\nu_f$ fixed at $0.3$ for both the zero-field scenario (black curve) and the non-zero constant external field one ($h = 0.5$, green curve). The impact of the phase diagram is evident from the behavior of both curves: for the green curve we can see that initially the energy rapidly increases until $\nu_i + \nu_f$ reaches the dashed red QPT line in Fig. \ref{gamma_quench}, while in the region between the two QPT lines, the curve rises more gradually and finally, after crossing the second QPT line, the curve begins to decrease. For the black curve, we observe the analoguous behavior, but other than gaining slightly more energy, the QPT lines are shifted with respect to the previous case since the intersections between the 3D phase diagram and the plane $h = 0$ are different.

\subsection{Quench of $h$ at given $\gamma$ and $\delta$}
\begin{figure}[H]
    \centering
    \includegraphics[width=0.5\linewidth]{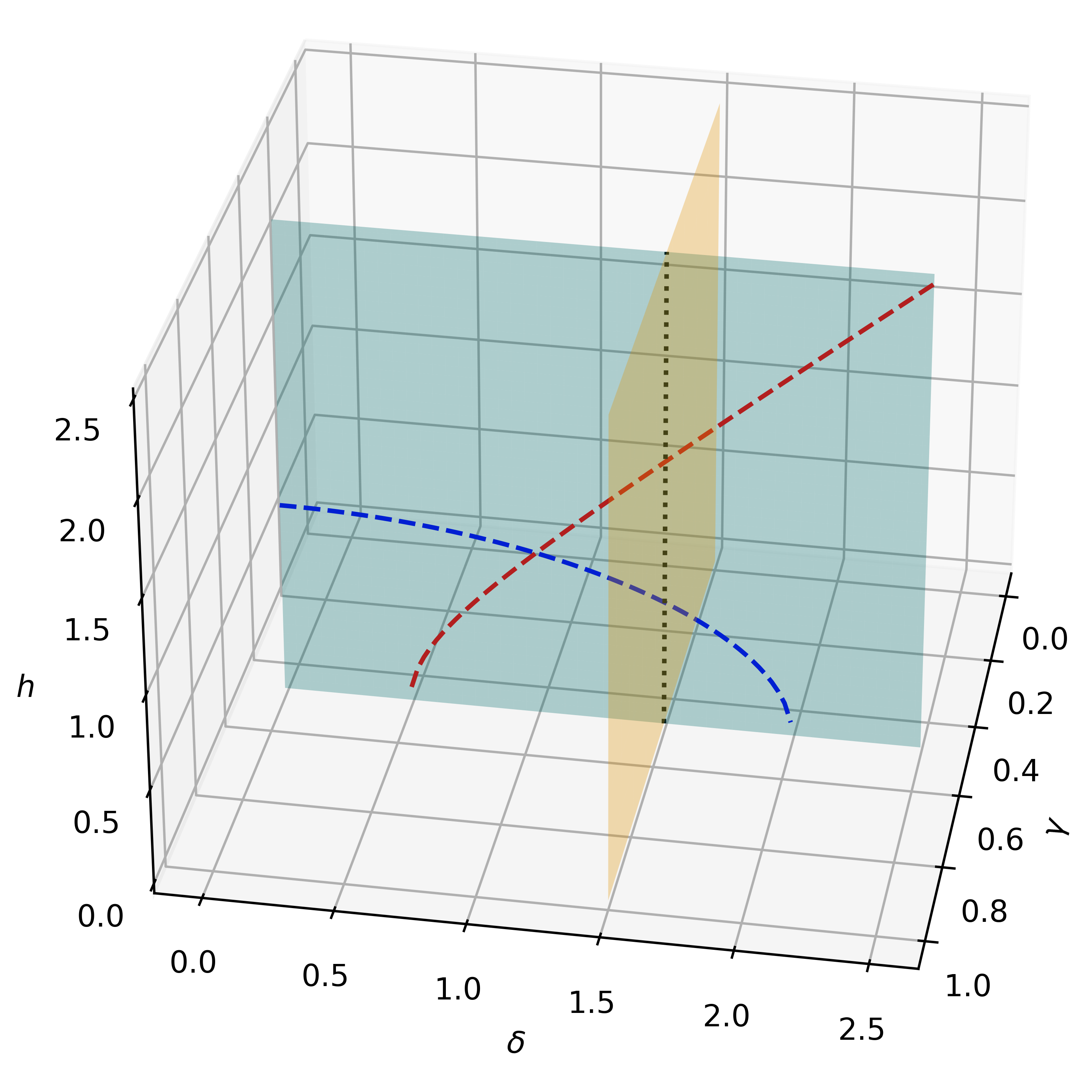}
    \caption{Section of the phase diagram of the model. The cyan and orange planes have equations $\gamma = 0.5$ and $\delta = 1.5$ respectively, while the black dotted line is given by the intersection of the aforementioned planes and it is the line where the quench takes place. The blue and red dashed lines are the QPT lines given by the intersection of the boundaries reported in Eq.\eqref{Hyp_Line} and Eq.\eqref{Cone_Line} with the $\gamma = 0.5$ plane.}
    \label{h_quench}
\end{figure}

\begin{figure}[H]
    \centering
    \includegraphics[width=0.6\linewidth]{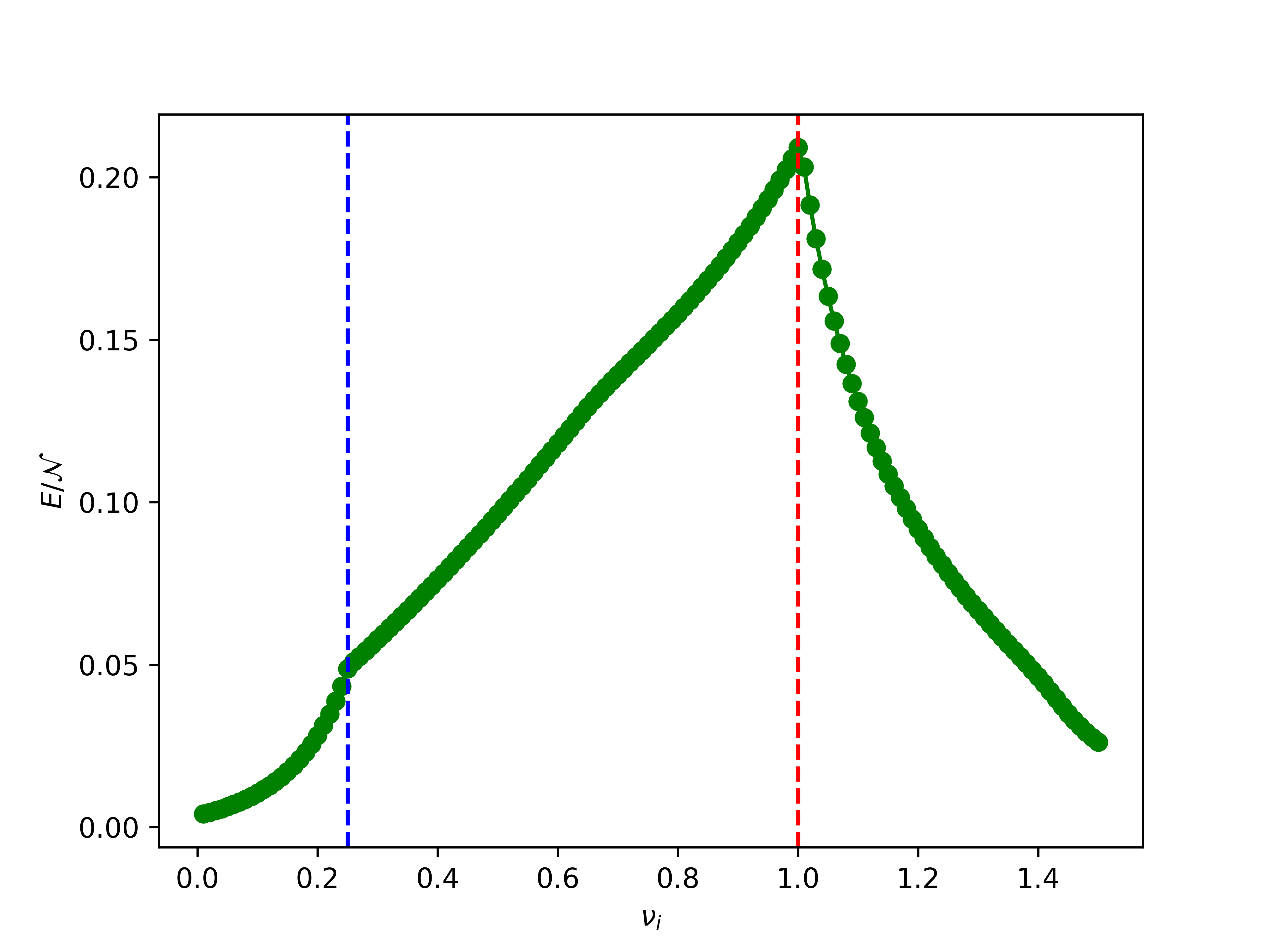}
    \caption{Energy stored per dimer as a function of $\nu_i$ after a quench of the external field at fixed $\nu_f = 0.41$, $\gamma = 0.5$ and $\delta = 1.5$ in the asymptotic regime, i.e. $\tau \to \infty$. The two dashed red and blue lines represent the values of $\nu_i$ such that the charging Hamiltonian depending on $\nu_i + \nu_f$ is critical.}
    \label{h_quench_RESULT}
\end{figure}

Now, we perform a quench of the external field, once again from an initial value $\nu_i$ to a final value $\nu_i + \nu_f$, along the line given by the intersection of the $\gamma = 0.5$ and the $\delta = 1.5$ planes, as shown in Fig. \ref{h_quench}.
As in the previous case, Figure \ref{h_quench_RESULT} shows the energy stored per dimer as a function of $\nu_i$, with $\nu_f$ fixed at $0.41$. The observed trend is very similar to the one shown in Fig.\ref{gamma_quench_RESULT}, with an initial rising until the first QPT line, an almost-linear rising until the second one and then a decrease.
\subsection{Quench of $h$ and $\gamma$ at given $\delta$}
\begin{figure}[H]
    \centering
    \includegraphics[width=0.5\linewidth]{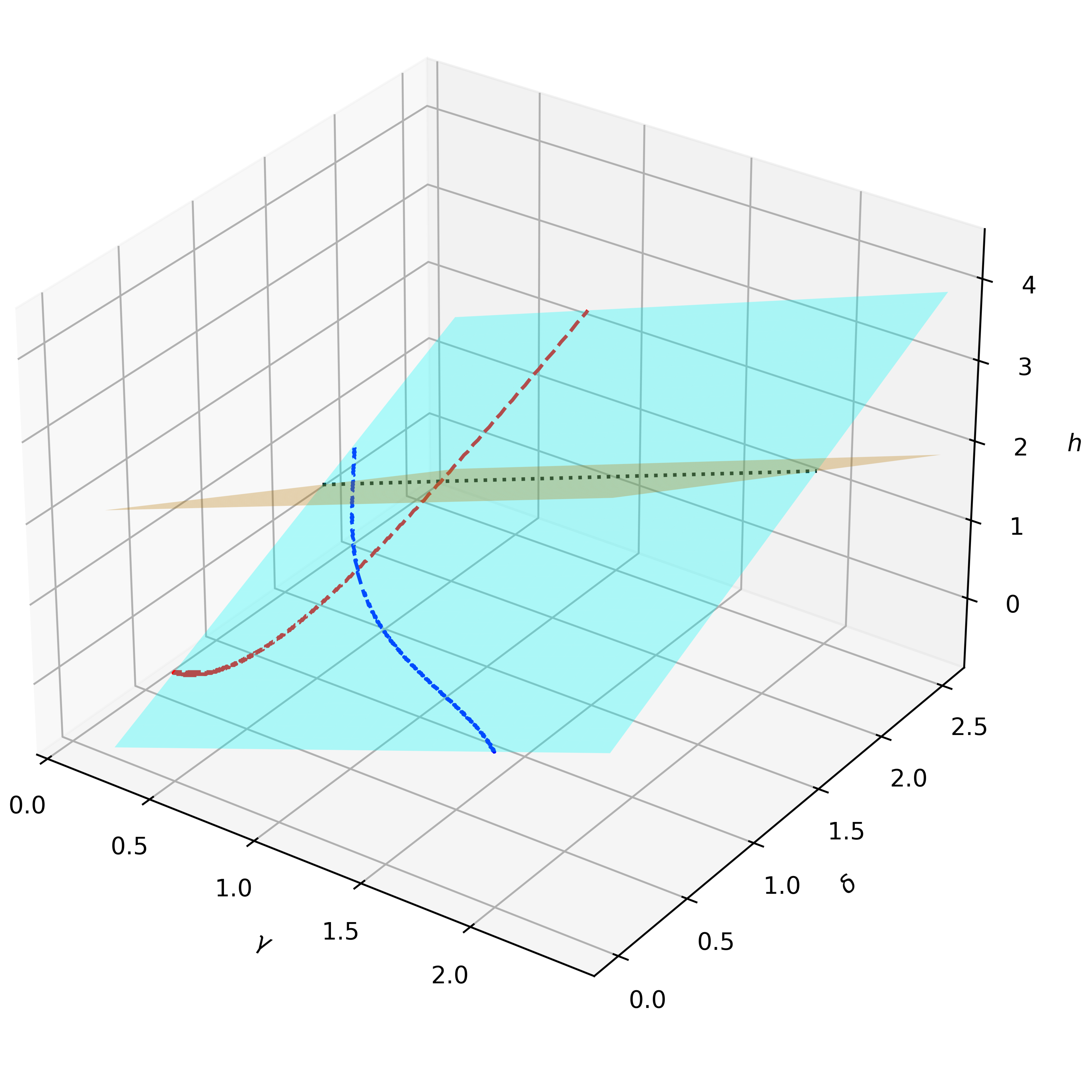}
    \caption{Section of the phase diagram of the model. The cyan and orange planes have equations $h = \gamma + \delta - 1$ and $h = \gamma - \delta + 2$ respectively, while the black dotted line is given by the intersection of the aforementioned planes and it is the line where the quench takes place. The blue and red dashed lines are the QPT lines given by the intersection of the boundaries reported in Eq.\eqref{Hyp_Line} and Eq.\eqref{Cone_Line} with the $h = \gamma + \delta - 1$ plane.}
    \label{h_gamma_quench}
\end{figure}

\begin{figure}[H]
    \centering
    \includegraphics[width=0.6\linewidth]{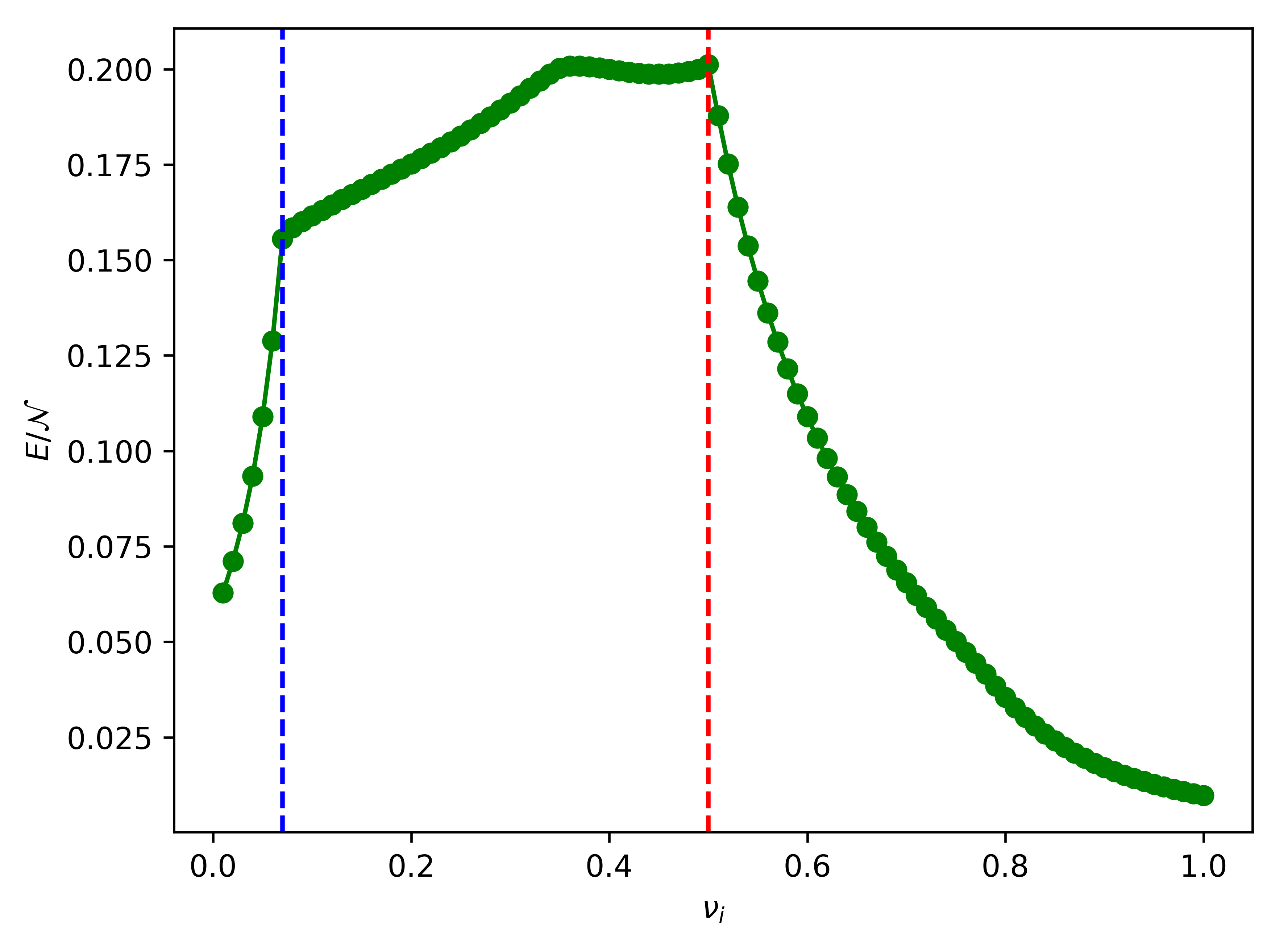}
    \caption{Energy stored per dimer as a function of $\nu_i$ after a quench of both the anisotropy parameter and the external field at fixed $\nu_f = 0.28$ and $\delta = 1.5$ in the asymptotic regime, i.e. $\tau \to \infty$. The two dashed red and blue lines represent the values of $\nu_i$ such that the charging Hamiltonian depending on $\nu_i + \nu_f$ is critical.}
    \label{h_gamma_quench_RESULT}
\end{figure}

Lastly, we vary both the anisotropy and the external field. The plane on which we perform the time evolution has the equation $h = \gamma + \delta - 1$ and the dotted black line has the following parametric equation
\begin{equation}
    \begin{cases}
        &\gamma = \nu_i \\
        &\delta = 1.5 \\
        &h = \nu_i + 0.5
    \end{cases}
\end{equation}
so that the quench of $\nu_i$ has an impact on both $\gamma$ and $h$. The QPT lines are represented in Fig. \ref{h_gamma_quench}
After fixing $\nu_f = 0.28$, we obtain the plot reported in Figure \ref{h_gamma_quench_RESULT}: even in this more complex scenario, we can observe the same features also presented in the previous two cases. 

\section{Conclusions}
In this work, we have investigated the interplay between QPTs and the energy storage properties of a dimerized XY spin chain subjected to an external transverse field, used as a QB. By extending previous analyses carried out in the absence of an external field, we have demonstrated that the introduction of a non-zero field enriches the quantum phase diagram and significantly influences the charging process. We considered three different scenarios: a quench of the anisotropy parameter $\gamma$, a quench of the external field $h$ and a simultaneous quench of both $\gamma$ and $h$. In all cases, we focused on the asymptotic regime of long charging times ($\tau \to \infty$), where the energy stored per dimer was analyzed as a function of the pre-quench parameter.

Our results reveal common patterns across all scenarios: first, the stored energy initially rises sharply until the first QPT line is crossed. In the region between two QPT lines, where a plateau was observed in the zero-field case ($h=0$) quenching the dimerization parameter \cite{Grazi24}, the energy now increases: this change reflects the role of both the type of quench and the external field in modifying the phenomenology of the plateau regime and suggests that QPT signatures at the level of the stored energy persist even for more complex models. Finally, after crossing the second QPT line, the stored energy begins to decrease. This behavior highlights the sensitivity of the system to the quantum phase diagram of the charging Hamiltonian. As an extension of this work, in order to have a more general understanding of the role of QPTs on quantum batteries based on integrable systems, it would be interesting to analyze scenarios in which the $\mathcal{M}$ matrix used to compute the energy stored in Eq.\eqref{DeltaE} can be accessed analitically. For instance, by analyzing the so-called cluster Ising model \cite{Smacchia11, Ding19}, one could assess the role of multi-spin interactions, while considering free fermions on generic lattices the effects related to dimensionality and topology might be unvailed \cite{Susskind1977, SHAMIR199390}. Additionally, the role of non-sudden quenches, initial thermal states, noise and dissipation needs to be addressed to translate our analysis to real-world quantum batteries \cite{Neyenhuis17, Knight89, Puskarov16, Cole13, Takei18, Gangadharaiah21, Troyer05, HongGang15, Giedke13, Katsura19}

\subsection*{Fundings}
N.T.Z. acknowledges the funding through the NextGenerationEu Curiosity Driven
Project "Understanding even-odd criticality". N.T.Z. acknowledge the funding
through the "Non-reciprocal supercurrent and topological transitions in hybrid Nb- InSb
nanoflags" project (Prot. 2022PH852L) in the framework of PRIN 2022 initiative of the
Italian Ministry of University (MUR) for the National Research Program (PNR). This project has been funded within the programme ``PNRR Missione 4 - Componente 2 - Investimento 1.1 Fondo per il Programma Nazionale di Ricerca e Progetti di Rilevante Interesse Nazionale (PRIN)''

D.F. acknowledges the contribution
of the European Union-NextGenerationEU through the
"Quantum Busses for Coherent Energy Transfer" (QUBERT) project, in the framework of the Curiosity Driven
2021 initiative of the University of Genova and through
the "Solid State Quantum Batteries: Characterization
and Optimization" (SoS-QuBa) project (Prot. 2022XK5CPX), in the framework of the PRIN 2022 initiative of the Italian Ministry
of University (MUR) for the National Research Program
(PNR). This project has been funded within the program ``PNRR Missione 4 - Componente 2 - Investimento 1.1 Fondo per il Programma Nazionale di Ricerca e Progetti di Rilevante Interesse Nazionale (PRIN)''.


\begin{thebibliography}{99}
\bibitem[1]{De16}
		De~las Cuevas, G.; Cubitt, T.S.
		\newblock Simple universal models capture all classical spin physics.
		\newblock \href{https://doi.org/10.1126/science.aab3326}{\newblock{\em Science} {\bf 2016}, {\em 351},~1180--1183}.
		
		\bibitem[2]{Auerbach12}
		Auerbach, A.
		\newblock {\em Interacting Electrons and Quantum Magnetism}; Springer Science 
		\& Business Media: Berlin/Heidelberg, Germany,
		2012.
		
		\bibitem[3]{Wysin15}
		Wysin, G.M.
		\newblock Magnetism theory: spin models. In {\em Magnetic Excitations and 
		Geometric Confinement}; 2053--2563
		;
		\newblock \href{https://doi.org/10.1088/978-0-7503-1074-1ch2}{IOP Publishing: Bristol, UK, 2015; pp. 2--1 to 2--47}.
		
		\bibitem[4]{cialone2017tailoring}
		Cialone, M.; Celegato, F.; Co{\"\i}sson, M.; Barrera, G.; Fiore, G.; Shvab, R.; Klement, 
		U.; Rizzi, P.; Tiberto, P.
		\newblock Tailoring magnetic properties of multicomponent layered structure via 
		current annealing in FePd thin films.
		\newblock \href{https://doi.org/10.1038/s41598-017-16963-5}{{\em Sci. Rep.} {\bf 2017}, {\em 7},~16691}.
		
		\bibitem[5]{cialone2020comparative}
		Cialone, M.; Fernandez-Barcia, M.; Celegato, F.; Coisson, M.; Barrera, G.; Uhlemann, 
		M.; Gebert, A.; Sort, J.; Pellicer, E.; Rizzi, P.;  et~al.
		\newblock A comparative study of the influence of the deposition technique 
		(electrodeposition versus sputtering) on the properties of nanostructured Fe70Pd30 
		films.
		\newblock \href{https://doi.org/10.1080/14686996.2020.1780097}{{\em Sci. Technol. Adv. Mater.} {\bf 2020}, {\em 21},~424--434}.
		
		\bibitem[6]{Manousakis91}
		Manousakis, E.
		\newblock The spin-\textonehalf{} Heisenberg antiferromagnet on a square lattice 
		and its application to the cuprous oxides.
		\newblock \href{https://doi.org/10.1103/RevModPhys.63.1}{{\em Rev. Mod. Phys.} {\bf 1991}, {\em 63},~1--62}.
		
		\bibitem[7]{Bramwell98}
		Bramwell, S.; Harris, M.
		\newblock Frustration in Ising-type spin models on the pyrochlore lattice.
		\newblock \href{https://doi.org/10.1088/0953-8984/10/14/002}{{\em J. Physics: Condens. Matter.} {\bf 1998}, {\em 10},~L215}.
		
		\bibitem[8]{Mydosh_2015}
		Mydosh, J.A.
		\newblock Spin glasses: redux: an updated experimental/materials survey.
		\newblock \href{https://doi.org/10.1088/0034-4885/78/5/052501}{{\em Rep. Prog. Phys.} {\bf 2015}, {\em 78},~052501}.
		
		\bibitem[9]{Pan17}
		Pan, F.; Chico, J.; Delin, A.; Bergman, A.; Bergqvist, L.
		\newblock Extended spin model in atomistic simulations of alloys.
		\newblock \href{https://doi.org/10.1103/PhysRevB.95.184432}{{\em Phys. Rev. B} {\bf 2017}, {\em 95},~184432}.
		
		\bibitem[10]{Cavaliere2023}
		Cavaliere, F.; Razzoli, L.; Carrega, M.; Benenti, G.; Sassetti, M.
		\newblock Hybrid quantum thermal machines with dynamical couplings.
		\newblock \href{https://doi.org/10.1016/j.isci.2023.106235}{{\em iScience} {\bf 2023}, {\em 26}, 106235}.
		
		\bibitem[11]{Eckhardt2022}
		Eckhardt, C.J.; Passetti, G.; Othman, M.; Karrasch, C.; Cavaliere, F.; Sentef, M.A.; 
		Kennes, D.M.
		\newblock Quantum Floquet engineering with an exactly solvable tight-binding chain 
		in a cavity.
		\newblock \href{https://doi.org/10.1038/s42005-022-00880-9}{{\em Commun. Phys.} {\bf 2022}, {\em 5},~122}.
		
		\bibitem[12]{XiaoYong07}
		Feng, X.Y.; Zhang, G.M.; Xiang, T.
		\newblock Topological Characterization of Quantum Phase Transitions in a 
		Spin-$1/2$ Model.
		\newblock \href{https://doi.org/10.1103/PhysRevLett.98.087204}{{\em Phys. Rev. Lett.} {\bf 2007}, {\em 98},~087204}.
		
		\bibitem[13]{Duivenvoorden13}
		{Duivenvoorden, K.; Quella, T.
			\newblock Topological phases of spin chains.
			\newblock \href{https://doi.org/10.1103/PhysRevB.87.125145}{{\em Phys. Rev. B} {\bf 2013}, {\em 87},~125145}.
		
		\bibitem[14]{Dubinkin19}
		Dubinkin, O.; Hughes, T.L.
		\newblock Higher-order bosonic topological phases in spin models.
		\newblock \href{https://doi.org/10.1103/PhysRevB.99.235132}{{\em Phys. Rev. B} {\bf 2019}, {\em 99},~235132}.
		
		\bibitem[15]{Balents10}
		Balents, L.
		\newblock Spin liquids in frustrated magnets.
		\newblock \href{https://doi.org/10.1038/nature08917}{{\em Nature} {\bf 2010}, {\em 464},~199--208}.
		
		\bibitem[16]{Diep13}
		Diep, H.T.
		\newblock {\em Frustrated Spin Systems}; World Scientific: Singapore, 2013.
		
		\bibitem[17]{Sacco24}
		Sacco~Shaikh, D.; Catalano, A.G.; Cavaliere, F.; Franchini, F.; Sassetti, M.; 
		Traverso~Ziani, N.
		\newblock Phase diagram of the topologically frustrated XY chain.
        \newblock
        \href{https://link.springer.com/article/10.1140/epjp/s13360-024-05534-z}{{\em  Eur. Phys. J. Plus} {\bf 2024}, {\em 139},~1--14}
		
		\bibitem[18]{maric2020quantum}
		Mari{\'c}, V.; Giampaolo, S.M.; Franchini, F.
		\newblock Quantum phase transition induced by topological frustration.
		\newblock \href{https://doi.org/10.1038/s42005-020-00486-z}{{\em Commun. Phys.} {\bf 2020}, {\em 3},~220}.
		
		\bibitem[19]{lacroix2011introduction}
		Lacroix, C.; Mendels, P.; Mila, F.
		\newblock {\em Introduction to Frustrated Magnetism: Materials, Experiments, 
		Theory};  Springer Science \& Business Media: Berlin/Heidelberg, Germany,
		2011; Volume 164,
		
		\bibitem[20]{Mitra18}
		Mitra, A.
		\newblock Quantum quench dynamics.
		\newblock \href{https://doi.org/10.1146/annurev-conmatphys-031016-025451}{{\em Annu. Rev. Condens. Matter Phys.} {\bf 2018}, {\em 9},~245--259}.
		
		\bibitem[21]{Essler16}
		Essler, F.H.; Fagotti, M.
		\newblock Quench dynamics and relaxation in isolated integrable quantum spin 
		chains.
		\newblock \href{https://doi.org/10.1088/1742-5468/2016/06/064002}{{\em J. Stat. Mech. Theory Exp.} {\bf 2016}, {\em 2016},~064002}.
		
		\bibitem[22]{Porta18}
		Porta, S.; Gambetta, F.M.; Traverso~Ziani, N.; Kennes, D.M.; Sassetti, M.; Cavaliere, F.
		\newblock Nonmonotonic response and light-cone freezing in fermionic systems 
		under quantum quenches from gapless to gapped or partially gapped states.
		\newblock \href{https://doi.org/10.1103/PhysRevB.97.035433}{{\em Phys. Rev. B} {\bf 2018}, {\em 97},~035433}.
		
		\bibitem[23]{Porta20}
		Porta, S.; Cavaliere, F.; Sassetti, M.; Traverso~Ziani, N.
		\newblock Topological classification of dynamical quantum phase transitions in the 
		xy chain.
		\newblock \href{https://www.nature.com/articles/s41598-020-69621-8}{{\em Sci. Rep.} {\bf 2020}, {\em 10},~12766}.
		
		\bibitem[24]{Faure19}
		Faure, Q.; Takayoshi, S.; Simonet, V.; Grenier, B.; M\aa{}nsson, M.; White, J.S.; Tucker, 
		G.S.; R\"uegg, C.; Lejay, P.; Giamarchi, T.;  et~al.
		\newblock Tomonaga-Luttinger Liquid Spin Dynamics in the 
		Quasi-One-Dimensional Ising-Like Antiferromagnet 
		${\mathrm{BaCo}}_{2}{\mathrm{V}}_{2}{\mathrm{O}}_{8}$.
		\newblock \href{https://doi.org/10.1103/PhysRevLett.123.027204}{{\em Phys. Rev. Lett.} {\bf 2019}, {\em 123},~027204}.
		
		\bibitem[25]{faure2018topological}
		Faure, Q.; Takayoshi, S.; Petit, S.; Simonet, V.; Raymond, S.; Regnault, L.P.; Boehm, M.; 
		White, J.S.; M{\aa}nsson, M.; R{\"u}egg, C.;  et~al.
		\newblock Topological quantum phase transition in the Ising-like antiferromagnetic 
		spin chain BaCo2V2O8.
		\newblock \href{https://doi.org/10.1038/s41567-018-0126-8}{{\em Nat. Phys.} {\bf 2018}, {\em 14},~716--722}.
		
		\bibitem[26]{kinoshita2006quantum}
		Kinoshita, T.; Wenger, T.; Weiss, D.S.
		\newblock A quantum Newton's cradle.
		\newblock
        \href{https://doi.org/10.1038/nature04693}{{\em Nature} {\bf 2006}, {\em 440},~900--903}.
		
		\bibitem[27]{Holstein40}
		Holstein, T.; Primakoff, H.
		\newblock Field Dependence of the Intrinsic Domain Magnetization of a Ferromagnet.
		\newblock \href{https://doi.org/10.1103/PhysRev.58.1098}{{\em Phys. Rev.} {\bf 1940}, {\em 58},~1098--1113}.
		
		\bibitem[28]{popov1988functional}
		Popov, V.N.; Fedotov, S.
		\newblock The functional-integration method and diagram technique for spin 
		systems.
		\newblock \href{https://www.semanticscholar.org/paper/The-functional-integration-method-and-diagram-for/ca4a2fea661c6f7b037a8549c85b0dc722d8e4ff}{{\em Zh. Eksp. Teor. Fiz} {\bf 1988}, {\em 94},~183--194}.
		
		\bibitem[29]{Traverso23clock}
		Traverso, S.; Fleckenstein, C.; Sassetti, M.; Ziani, N.T.
		\newblock {An exact local mapping from clock-spins to fermions}.
		\newblock \href{https://doi.org/10.21468/SciPostPhysCore.6.3.055}{{\em SciPost Phys. Core} {\bf 2023}, {\em 6},~055}.
		
		\bibitem[30]{Ziani17}
		Traverso~Ziani, N.; Fleckenstein, C.; Dolcetto, G.; Trauzettel, B.
		\newblock Fractional charge oscillations in quantum dots with quantum spin Hall 
		effect.
		\newblock \href{https://doi.org/10.1103/PhysRevB.95.205418}{{\em Phys. Rev. B} {\bf 2017}, {\em 95},~205418}.
		
		\bibitem[31]{rodriguez2020relaxation}
		Rodriguez, R.; Parmentier, F.; Ferraro, D.; Roulleau, P.; Gennser, U.; Cavanna, A.; 
		Sassetti, M.; Portier, F.; Mailly, D.; Roche, P.
		\newblock Relaxation and revival of quasiparticles injected in an interacting quantum 
		Hall liquid.
		\newblock \href{https://doi.org/10.1038/s41467-020-16331-4}{{\em Nat. Commun.} {\bf 2020}, {\em 11},~2426}.
		
		\bibitem[32]{Gambetta15}
		Gambetta, F.M.; Ziani, N.T.; Barbarino, S.; Cavaliere, F.; Sassetti, M.
		\newblock Anomalous Friedel oscillations in a quasihelical quantum dot.
		\newblock \href{https://doi.org/10.1103/PhysRevB.91.235421}{{\em Phys. Rev. B} {\bf 2015}, {\em 91},~235421}.
		
		\bibitem[33]{TraversoZiani_2013}
		Ziani, N.T.; Cavaliere, F.; Sassetti, M.
		\newblock Theory of the STM detection of Wigner molecules in spin-incoherent 
		CNTs.
		\newblock \href{https://doi.org/10.1209/0295-5075/102/47006}{{\em Europhysics Letters} {\bf 2013}, {\em 102},~47006}.
		
		\bibitem[34]{cryst11010020}
		Ziani, N.T.; Cavaliere, F.; Becerra, K.G.; Sassetti, M.
		\newblock A Short Review of One-Dimensional Wigner Crystallization.
		\newblock \href{https://doi.org/10.3390/cryst11010020}{{\em Crystals} {\bf 2021}, {\em 11}}.
		
		\bibitem[35]{Bloch08}
		Bloch, I.; Dalibard, J.; Zwerger, W.
		\newblock Many-body physics with ultracold gases.
		\newblock \href{https://doi.org/10.1103/RevModPhys.80.885}{{\em Rev. Mod. Phys.} {\bf 2008}, {\em 80},~885--964}.
		
		\bibitem[36]{Cazalilla11}
		Cazalilla, M.A.; Citro, R.; Giamarchi, T.; Orignac, E.; Rigol, M.
		\newblock One dimensional bosons: From condensed matter systems to ultracold 
		gases.
		\newblock \href{https://doi.org/10.1103/RevModPhys.83.1405}{{\em Rev. Mod. Phys.} {\bf 2011}, {\em 83},~1405--1466}.
		
		\bibitem[37]{trotzky2012probing}
		Trotzky, S.; Chen, Y.A.; Flesch, A.; McCulloch, I.P.; Schollw{\"o}ck, U.; Eisert, J.; Bloch, 
		I.
		\newblock Probing the relaxation towards equilibrium in an isolated strongly 
		correlated one-dimensional Bose gas.
		\newblock \href{https://doi.org/10.1038/nphys2232}{{\em Nat. Phys.} {\bf 2012}, {\em 8},~325--330}.
		
		\bibitem[38]{Imambekov09}
		Imambekov, A.; Glazman, L.I.
		\newblock Universal Theory of Nonlinear Luttinger Liquids.
		\newblock \href{https://doi.org/10.1126/science.1165403}{{\em Science} {\bf 2009}, {\em 323},~228--231}.
		
		\bibitem[39]{Imambekov12}
		Imambekov, A.; Schmidt, T.L.; Glazman, L.I.
		\newblock One-dimensional quantum liquids: Beyond the Luttinger liquid paradigm.
		\newblock \href{https://doi.org/10.1103/RevModPhys.84.1253}{{\em Rev. Mod. Phys.} {\bf 2012}, {\em 84},~1253--1306}.
		
		\bibitem[40]{Wu06}
		Wu, C.; Bernevig, B.A.; Zhang, S.C.
		\newblock Helical Liquid and the Edge of Quantum Spin Hall Systems.
		\newblock \href{https://doi.org/10.1103/PhysRevLett.96.106401}{{\em Phys. Rev. Lett.} {\bf 2006}, {\em 96},~106401}.
		
		\bibitem[41]{Fiete06}
		Fiete, G.A.; Le~Hur, K.; Balents, L.
		\newblock Coulomb drag between two spin-incoherent Luttinger liquids.
		\newblock \href{https://doi.org/10.1103/PhysRevB.73.165104}{{\em Phys. Rev. B} {\bf 2006}, {\em 73},~165104}.
		
		\bibitem[42]{Fiete07}
		Fiete, G.A.
		\newblock Colloquium: The spin-incoherent Luttinger liquid.
		\newblock \href{https://doi.org/10.1103/RevModPhys.79.801}{{\em Rev. Mod. Phys.} {\bf 2007}, {\em 79},~801--820}.
		
		\bibitem[43]{Matveev07}
		Matveev, K.A.; Furusaki, A.; Glazman, L.I.
		\newblock Bosonization of strongly interacting one-dimensional electrons.
		\newblock \href{https://doi.org/10.1103/PhysRevB.76.155440}{{\em Phys. Rev. B} {\bf 2007}, {\em 76},~155440}.
		
		\bibitem[44]{JVoit_1995}
		Voit, J.
		\newblock One-dimensional Fermi liquids.
		\newblock \href{https://doi.org/10.1088/0034-4885/58/9/002}{{\em Rep. Prog. Phys.} {\bf 1995}, {\em 58},~977}.
		
		\bibitem[45]{Haldane81}
		Haldane, F.D.M.
		\newblock Effective Harmonic-Fluid Approach to Low-Energy Properties of 
		One-Dimensional Quantum Fluids.
		\newblock \href{https://doi.org/10.1103/PhysRevLett.47.1840}{{\em Phys. Rev. Lett.} {\bf 1981}, {\em 47},~1840--1843}.
		
		\bibitem[46]{Haldane_1981}
		Haldane, F.D.M.
		\newblock 'Luttinger liquid theory' of one-dimensional quantum fluids. I. Properties 
		of the Luttinger model and their extension to the general 1D interacting spinless Fermi 
		gas.
		\newblock \href{https://doi.org/10.1088/0022-3719/14/19/010}{{\em J. Phys. Solid State Phys.} {\bf 1981}, {\em 14},~2585}.
		
		\bibitem[47]{giamarchibook}
		Giamarchi, T.
		\newblock {\em Quantum Physics in One Dimension}; Oxford University Press: 
		Oxford, UK,  2003.
		
		\bibitem[48]{bethe1931theorie}
		Bethe, H.
		\newblock Zur theorie der metalle: I. Eigenwerte und eigenfunktionen der linearen 
		atomkette.
		\newblock \href{https://doi.org/10.1007/BF01341708}{{\em Zeitschrift f{\"u}r Physik} {\bf 1931}, {\em 71},~205--226}.
		
		\bibitem[49]{Jordan28}
		Jordan, P.; Wigner, E.
		\newblock {\"U}ber das Paulische {\"A}quivalenzverbot.
		\newblock \href{https://doi.org/10.1007/BF01331938}{{\em Z. Phys.} {\bf 1928}, {\em 47},~631--651}.
		
		\bibitem[50]{Franchini17}
		Franchini, F.
		\newblock {\em An Introduction to Integrable Techniques for One-Dimensional 
		Quantum Systems};  Springer: Berlin/Heidelberg, Germany,
		2017; Volume 940.
		
		\bibitem[51]{Bayat22}
		Bayat, A.; Bose, S.; Johannesson, H.
		\newblock {\em Entanglement in Spin Chains: From Theory to Quantum Technology 
		Applications}; Springer: Berlin/Heidelberg, Germany,
		2022.
		
		\bibitem[52]{Le_2018}
		Le, T.P.; Levinsen, J.; Modi, K.; Parish, M.M.; Pollock, F.A.
		\newblock Spin-chain model of a many-body quantum battery.
		\newblock \href{https://doi.org/10.1103/PhysRevA.97.022106}{{\em Phys. Rev. A} {\bf 2018}, {\em 97},~022106}.
		
		\bibitem[53]{Liu21}
		Liu, J.X.; Shi, H.L.; Shi, Y.H.; Wang, X.H.; Yang, W.L.
		\newblock Entanglement and work extraction in the central-spin quantum battery.
		\newblock \href{https://doi.org/10.1103/PhysRevB.104.245418}{{\em Phys. Rev. B} {\bf 2021}, {\em 104},~245418}.
		
		\bibitem[54]{Zhao21}
		Zhao, F.; Dou, F.Q.; Zhao, Q.
		\newblock Quantum battery of interacting spins with environmental noise.
		\newblock \href{https://doi.org/10.1103/PhysRevA.103.033715}{{\em Phys. Rev. A} {\bf 2021}, {\em 103},~033715}.
		
		\bibitem[55]{Catalano23}
		Catalano, A.; Giampaolo, S.; Morsch, O.; Giovannetti, V.; Franchini, F.
		\newblock Frustrating Quantum Batteries.
		\newblock \href{https://doi.org/10.1103/PRXQuantum.5.030319}{{\em PRX Quantum} {\bf 2024}, {\em 5},~030319}.
		
		\bibitem[56]{Grazi24}
		Grazi, R.; Sacco~Shaikh, D.; Sassetti, M.; Traverso~Ziani, N.; Ferraro, D.
		\newblock Controlling Energy Storage Crossing Quantum Phase Transitions in an 
		Integrable Spin Quantum Battery.
		\newblock \href{https://doi.org/10.1103/PhysRevLett.133.197001}{{\em Phys. Rev. Lett.} {\bf 2024}, {\em 133},~197001}.
		
		\bibitem[57]{Ali24SuperExt}
		Ali, A.; Elghaayda, S.; Al-Kuwari, S.; Hussain, M.I.; Rahim, M.T.; Kuniyil, H.; Seuda, 
		C.; Allati, A.E.; Mansour, M.; Haddadi, S.
		\newblock Super-Extensive Scaling in 1D Spin$-1/2$ $XY-\Gamma(\gamma)$ Chain 
		Quantum Battery.
		\emph{arXiv}  \textbf{2024},  \href{https://doi.org/10.48550/arXiv.2411.14074}{arXiv:2411.14074}.
		
		\bibitem[58]{Alicki_2013}
		Alicki, R.; Fannes, M.
		\newblock Entanglement boost for extractable work from ensembles of quantum 
		batteries. \newblock
        \href{https://doi.org/10.1103/PhysRevE.87.042123}{{\em Phys. Rev. E} {\bf 2013}, {\em 87},~042123}.
		
		\bibitem[59]{Bhattacharjee21}
		Bhattacharjee, S.; Dutta, A.
		\newblock Quantum thermal machines and batteries.
		\newblock \href{https://doi.org/10.1140/epjb/s10051-021-00235-3}{{\em  Eur. Phys. J. } {\bf 2021}, {\em 94}}.
		
		\bibitem[60]{Campaioli23}
		Campaioli, F.; Gherardini, S.; Quach, J.Q.; Polini, M.; Andolina, G.M.
		\newblock Colloquium: Quantum batteries.
		\newblock \href{https://doi.org/10.1103/RevModPhys.96.031001}{{\em Rev. Mod. Phys.} {\bf 2024}, {\em 96},~031001}.
		
		\bibitem[61]{Quach23}
		Quach, J.; Cerullo, G.; Virgili, T.
		\newblock Quantum batteries: The future of energy storage?
		\newblock \href{https://doi.org/https://doi.org/10.1016/j.joule.2023.09.003}{{\em Joule} {\bf 2023}, {\em 7},~2195--2200}.
		
		\bibitem[62]{Benenti17}
		Benenti, G.; Casati, G.; Saito, K.; Whitney, R.S.
		\newblock Fundamental aspects of steady-state conversion of heat to work at the 
		nanoscale.
		\newblock Fundamental aspects of steady-state conversion of heat to work at the 
		nanoscale.
        \newblock \href{https://doi.org/https://doi.org/10.1016/j.physrep.2017.05.008}{{\em Phys. Rep.} {\bf 2017}, {\em 694},~1--124}.
		
		\bibitem[63]{Esposito09}
		Esposito, M.; Harbola, U.; Mukamel, S.
		\newblock Nonequilibrium fluctuations, fluctuation theorems, and counting statistics 
		in quantum systems.
		\newblock \href{https://doi.org/10.1103/RevModPhys.81.1665}{{\em Rev. Mod. Phys.} {\bf 2009}, {\em 81},~1665--1702}.
		
		\bibitem[64]{Campisi16}
		Campisi, M.; Fazio, R.
		\newblock Dissipation, correlation and lags in heat engines.
		\newblock \href{https://doi.org/10.1088/1751-8113/49/34/345002}{{\em J. Phys. Math. Theor.} {\bf 2016}, {\em 49},~345002}.
		
		\bibitem[65]{Vinjanampathy16}
		Vinjanampathy, S.; Anders, J.
		\newblock Quantum thermodynamics.
		\newblock 
		\href{https://doi.org/10.1080/00107514.2016.1201896}{{\em Contemp. Phys.} {\bf 2016}, {\em 57},~545--579}.
		
		\bibitem[66]{Potts24}
		Potts, P.P.
		\newblock Quantum Thermodynamics. \newblock \href{https://doi.org/10.48550/arXiv.2406.19206}{\emph{arXiv}  \textbf{2024},  
		arXiv:2406.19206}.
		
		
		\bibitem[67]{Hu_2022}
		Hu, C.K.; Qiu, J.; Souza, P.J.P.; Yuan, J.; Zhou, Y.; Zhang, L.; Chu, J.; Pan, X.; Hu, L.; Li, 
		J.;  et~al.
		\newblock Optimal charging of a superconducting quantum battery.
		\newblock \href{https://doi.org/10.1088/2058-9565/ac8444}{{\em Quantum Sci. Technol.} {\bf 2022}, {\em 7},~045018}.
		
		\bibitem[68]{Gemme23}
		Gemme, G.; Andolina, G.M.; Pellegrino, F.M.D.; Sassetti, M.; Ferraro, D.
		\newblock Off-Resonant Dicke Quantum Battery: Charging by Virtual Photons.
		\newblock \href{https://doi.org/10.3390/batteries9040197}{{\em Batteries} {\bf 2023}, {\em 9}, 197}.
		
		\bibitem[69]{Dou23}
		Dou, F.Q.; Yang, F.M.
		\newblock Superconducting transmon qubit-resonator quantum battery.
		\newblock \href{https://doi.org/10.1103/PhysRevA.107.023725}{{\em Phys. Rev. A} {\bf 2023}, {\em 107},~023725}.
		
		\bibitem[70]{Razzoli24}
		Razzoli, L.; Gemme, G.; Khomchenko, I.; Sassetti, M.; Ouerdane, H.; Ferraro, D.; 
		Benenti, G.
		\newblock Cyclic solid-state quantum battery: Thermodynamic characterization and 
		quantum hardware simulation.    \newblock \href{https://doi.org/10.1088/2058-9565/ad9ed4}{\emph{Quantum Sci. Technol.} \textbf{2025}, 10(1), 015064}.
		
		\bibitem[71]{Cavaliere24}
		Cavaliere, F.; Gemme, G.; Benenti, G.; Ferraro, D.; Sassetti, M.
		\newblock Dynamical blockade of a reservoir for optimal performances of a 
		quantum battery.  \href{https://doi.org/10.48550/arXiv.2407.16471}{\emph{arXiv} \textbf{2024},  arXiv:2407.16471}.
		
		\bibitem[72]{Chiribella21}
		Chiribella, G.; Yang, Y.; Renner, R.
		\newblock Fundamental Energy Requirement of Reversible Quantum Operations.
		\newblock \href{https://doi.org/10.1103/PhysRevX.11.021014}{{\em Phys. Rev. X} {\bf 2021}, {\em 11},~021014}.
		
		\bibitem[73]{Menta24}
		Menta, R.; Cioni, F.; Aiudi, R.; Polini, M.; Giovannetti, V.
		\newblock Globally driven superconducting quantum computing architecture.  
		\emph{arXiv} \textbf{2024},  arXiv:2407.01182.
		
		
		\bibitem[74]{Elyasi2024}
		Elyasi, S.N.; Rossi, M.A.C.; Genoni, M.G.
		\newblock Experimental simulation of daemonic work extraction in open quantum 
		batteries on a digital quantum computer  \emph{arXiv}, \textbf{2024}, 
		arXiv:2410.16567.
		
		\bibitem[75]{Binder_2015}
		Binder, F.C.; Vinjanampathy, S.; Modi, K.; Goold, J.
		\newblock Quantacell: powerful charging of quantum batteries.
		\newblock \href{https://doi.org/10.1088/1367-2630/17/7/075015}{{\em New J. Phys.} {\bf 2015}, {\em 17},~075015}.
		
		\bibitem[76]{Andolina18}
		Andolina, G.M.; Farina, D.; Mari, A.; Pellegrini, V.; Giovannetti, V.; Polini, M.
		\newblock Charger-mediated energy transfer in exactly solvable models for quantum 
		batteries.
		\newblock \href{https://doi.org/10.1103/PhysRevB.98.205423}{{\em Phys. Rev. B} {\bf 2018}, {\em 98},~205423}.
		
		\bibitem[77]{Crescente22}
		Crescente, A.; Ferraro, D.; Carrega, M.; Sassetti, M.
		\newblock Enhancing coherent energy transfer between quantum devices via a 
		mediator.
		\newblock \href{https://doi.org/10.1103/PhysRevResearch.4.033216}{{\em Phys. Rev. Res.} {\bf 2022}, {\em 4},~033216}.
		
		\bibitem[78]{PERK1975319}
		Perk, J.; Capel, H.; Zuilhof, M.; Siskens, T.
		\newblock On a soluble model of an antiferromagnetic chain with alternating 
		interactions and magnetic moments.
		\newblock \href{https://doi.org/https://doi.org/10.1016/0378-4371(75)90052-7}{{\em Phys. Stat. Mech. Its Appl.} {\bf 1975}, {\em 81},~319--348}.
		
		\bibitem[79]{wakatsuki2014fermion}
		Wakatsuki, R.; Ezawa, M.; Tanaka, Y.; Nagaosa, N.
		\newblock Fermion fractionalization to Majorana fermions in a dimerized Kitaev 
		superconductor.
		\newblock \href{https://doi.org/10.1103/PhysRevB.90.014505}{{\em Phys. Rev. B} {\bf 2014}, {\em 90},~014505}.
		
		\bibitem[80]{Ziani20}
		Ziani, N.T.; Fleckenstein, C.; Vigliotti, L.; Trauzettel, B.; Sassetti, M.
		\newblock From fractional solitons to Majorana fermions in a paradigmatic model of 
		topological superconductivity.
		\newblock \href{https://doi.org/10.1103/PhysRevB.101.195303}{{\em Phys. Rev. B} {\bf 2020}, {\em 101},~195303}.
		
		\bibitem[81]{Jackiw76}
		Jackiw, R.; Rebbi, C.
		\newblock Solitons with fermion number \textonehalf{}.
		\newblock \href{https://doi.org/10.1103/PhysRevD.13.3398}{{\em Phys. Rev. D} {\bf 1976}, {\em 13},~3398--3409}.
		
		\bibitem[82]{Kivelson82}
		Kivelson, S.; Schrieffer, J.R.
		\newblock Fractional charge, a sharp quantum observable.
		\newblock \href{https://doi.org/10.1103/PhysRevB.25.6447}{{\em Phys. Rev. B} {\bf 1982}, {\em 25},~6447--6451}.
		
		\bibitem[83]{Goldstone81}
		Goldstone, J.; Wilczek, F.
		\newblock Fractional Quantum Numbers on Solitons.
		\newblock \href{https://doi.org/10.1103/PhysRevLett.47.986}{{\em Phys. Rev. Lett.} {\bf 1981}, {\em 47},~986--989}.
		
		\bibitem[84]{Heeger88}
		Heeger, A.J.; Kivelson, S.; Schrieffer, J.R.; Su, W.P.
		\newblock Solitons in conducting polymers.
		\newblock \href{https://doi.org/10.1103/RevModPhys.60.781}{{\em Rev. Mod. Phys.} {\bf 1988}, {\em 60},~781--850}.
		
		\bibitem[85]{qi2008fractional}
		Qi, X.L.; Hughes, T.L.; Zhang, S.C.
		\newblock Fractional charge and quantized current in the quantum spin Hall state.
		\newblock \href{https://doi.org/10.1038/nphys913}{{\em Nat. Phys.} {\bf 2008}, {\em 4},~273--276}.
		
		\bibitem[86]{Fleckenstein21}
		Fleckenstein, C.; Ziani, N.T.; Calzona, A.; Sassetti, M.; Trauzettel, B.
		\newblock Formation and detection of Majorana modes in quantum spin Hall 
		trenches.
		\newblock \href{https://doi.org/10.1103/PhysRevB.103.125303}{{\em Phys. Rev. B} {\bf 2021}, {\em 103},~125303}.
		
		\bibitem[87]{traverso2024emerging}
		Traverso, S.; Sassetti, M.; Traverso~Ziani, N.
		\newblock Emerging topological bound states in Haldane model zigzag nanoribbons.
		\newblock \href{https://doi.org/10.1038/s41535-023-00615-1}{{\em npj Quantum Mater.} {\bf 2024}, {\em 9},~9}.
		
		\bibitem[88]{Traverso22}
		Traverso, S.; Traverso~Ziani, N.; Sassetti, M.
		\newblock Effects of the Vertices on the Topological Bound States in a 
		Quasicrystalline Topological Insulator.
		\newblock \href{https://doi.org/10.3390/sym14081736}{{\em Symmetry} {\bf 2022}, {\em 14}}.
		
		\bibitem[89]{Traverso22role}
		Traverso, S.; Sassetti, M.; Ziani, N.T.
		\newblock Role of the edges in a quasicrystalline Haldane model.
		\newblock \href{https://doi.org/10.1103/PhysRevB.106.125428}{{\em Phys. Rev. B} {\bf 2022}, {\em 106},~125428}.
		
		\bibitem[90]{AYuKitaev_2001}
		Kitaev, A.Y.
		\newblock Unpaired Majorana fermions in quantum wires.
		\newblock \href{https://doi.org/10.1070/1063-7869/44/10S/S29}{{\em Physics-Uspekhi} {\bf 2001}, {\em 44},~131}.
		
		\bibitem[91]{Deng_2016}
		Deng, M.T.; Vaitiekėnas, S.; Hansen, E.B.; Danon, J.; Leijnse, M.; Flensberg, K.; 
		Nygård, J.; Krogstrup, P.; Marcus, C.M.
		\newblock Majorana bound state in a coupled quantum-dot hybrid-nanowire system.
		\newblock 
        \href{https://10.1126/science.aaf3961}{{\em Science} {\bf 2016}, {\em 354},~1557--1562}.
		
		\bibitem[92]{Prada_2018}
		Pe\~naranda, F.; Aguado, R.; San-Jose, P.; Prada, E.
		\newblock Quantifying wave-function overlaps in inhomogeneous Majorana 
		nanowires.
		\newblock \href{https://doi.org/10.1103/PhysRevB.98.235406}{{\em Phys. Rev. B} {\bf 2018}, {\em 98},~235406}.
		
		\bibitem[93]{Fleckenstein_2018}
		Fleckenstein, C.; Dom\'{\i}nguez, F.; Traverso~Ziani, N.; Trauzettel, B.
		\newblock Decaying spectral oscillations in a Majorana wire with finite coherence 
		length.
		\newblock \href{https://doi.org/10.1103/PhysRevB.97.155425}{{\em Phys. Rev. B} {\bf 2018}, {\em 97},~155425}.
		
		\bibitem[94]{Dibyendu_2013}
		Roy, D.; Bondyopadhaya, N.; Tewari, S.
		\newblock Topologically trivial zero-bias conductance peak in semiconductor 
		Majorana wires from boundary effects.
		\newblock \href{https://doi.org/10.1103/PhysRevB.88.020502}{{\em Phys. Rev. B} {\bf 2013}, {\em 88},~020502}.
		
		\bibitem[95]{Flensberg_2010}
		Flensberg, K.
		\newblock Tunneling characteristics of a chain of Majorana bound states.
		\newblock \href{https://doi.org/10.1103/PhysRevB.82.180516}{{\em Phys. Rev. B} {\bf 2010}, {\em 82},~180516}.
		
		\bibitem[96]{Law_2009}
		Law, K.T.; Lee, P.A.; Ng, T.K.
		\newblock Majorana Fermion Induced Resonant Andreev Reflection.
		\newblock \href{https://doi.org/10.1103/PhysRevLett.103.237001}{{\em Phys. Rev. Lett.} {\bf 2009}, {\em 103},~237001}.
		
		\bibitem[97]{Prada_2020}
		Prada, E.; San-Jose, P.; de~Moor, M.W.A.; Geresdi, A.; Lee, E.J.H.; Klinovaja, J.; Loss, 
		D.; Nyg{\aa}rd, J.; Aguado, R.; Kouwenhoven, L.P.
		\newblock From Andreev to Majorana bound states in hybrid 
		superconductor--semiconductor nanowires.
		\newblock \href{https://doi.org/10.1038/s42254-020-0228-y}{{\em Nat. Rev. Phys.} {\bf 2020}, {\em 2},~575--594}.
		
		\bibitem[98]{Ivanov_2001}
		Ivanov, D.A.
		\newblock Non-Abelian Statistics of Half-Quantum Vortices in ${p}$-Wave 
		Superconductors.
		\newblock \href{https://doi.org/10.1103/PhysRevLett.86.268}{{\em Phys. Rev. Lett.} {\bf 2001}, {\em 86},~268--271}.
		
		\bibitem[99]{Marra_2022}
		Marra, P.
		\newblock Majorana nanowires for topological quantum computation.
		\newblock
        \href{https://pubs.aip.org/aip/jap/article/132/23/231101/2837934/Majorana-nanowires-for-topological-quantum}{{\em J. Appl. Phys.} {\bf 2022}, {\em 132},~231101}.
		
		\bibitem[100]{Leijnse_2012}
		Leijnse, M.; Flensberg, K.
		\newblock Introduction to topological superconductivity and Majorana fermions.
		\newblock \href{https://doi.org/10.1088/0268-1242/27/12/124003}{{\em Semicond. Sci. Technol.} {\bf 2012}, {\em 27},~124003}.
		
		\bibitem[101]{Nayak_2008}
		Nayak, C.; Simon, S.H.; Stern, A.; Freedman, M.; Das~Sarma, S.
		\newblock Non-Abelian anyons and topological quantum computation.
		\newblock \href{https://doi.org/10.1103/RevModPhys.80.1083}{{\em Rev. Mod. Phys.} {\bf 2008}, {\em 80},~1083--1159}.
		
		\bibitem[102]{Kitaev_2003}
		Kitaev, A.
		\newblock Fault-tolerant quantum computation by anyons.
		\newblock \href{https://doi.org/https://doi.org/10.1016/S0003-4916(02)00018-0}{{\em Ann. Phys.} {\bf 2003}, {\em 303},~2--30}.
		
		\bibitem[103]{Oreg_2010}
		Oreg, Y.; Refael, G.; von Oppen, F.
		\newblock Helical Liquids and Majorana Bound States in Quantum Wires.
		\newblock \href{https://doi.org/10.1103/PhysRevLett.105.177002}{{\em Phys. Rev. Lett.} {\bf 2010}, {\em 105},~177002}.
		
		\bibitem[104]{Lutchyn_2010}
		Lutchyn, R.M.; Sau, J.D.; Das~Sarma, S.
		\newblock Majorana Fermions and a Topological Phase Transition in 
		Semiconductor-Superconductor Heterostructures.
		\newblock \href{https://doi.org/10.1103/PhysRevLett.105.077001}{{\em Phys. Rev. Lett.} {\bf 2010}, {\em 105},~077001}.
		
		\bibitem[105]{Smacchia11}
		Smacchia, P.; Amico, L.; Facchi, P.; Fazio, R.; Florio, G.; Pascazio, S.; Vedral, V.
		\newblock Statistical mechanics of the cluster Ising model.
		\newblock \href{https://doi.org/10.1103/PhysRevA.84.022304}{{\em Phys. Rev. A} {\bf 2011}, {\em 84},~022304}.
		
		\bibitem[106]{Ding19}
		Ding, C.
		\newblock Phase transitions of a cluster Ising model.
		\newblock \href{https://doi.org/10.1103/PhysRevE.100.042131}{{\em Phys. Rev. E} {\bf 2019}, {\em 100},~042131}.
		
		\bibitem[107]{Susskind1977}
		Susskind, L.
		\newblock Lattice fermions.
		\newblock \href{https://doi.org/10.1103/PhysRevD.16.3031}{{\em Phys. Rev. D} {\bf 1977}, {\em 16},~3031--3039}.
		
		\bibitem[108]{SHAMIR199390}
		Shamir, Y.
		\newblock Chiral fermions from lattice boundaries.
		\newblock \href{https://doi.org/https://doi.org/10.1016/0550-3213(93)90162-I}{{\em Nucl. Phys. } {\bf 1993}, {\em 406},~90--106}.
		
		\bibitem[109]{Neyenhuis17}
		Neyenhuis, B.; Zhang, J.; Hess, P.W.; Smith, J.; Lee, A.C.; Richerme, P.; Gong, Z.X.; 
		Gorshkov, A.V.; Monroe, C.
		\newblock Observation of prethermalization in long-range interacting spin chains.
		\newblock
        \href{https://10.1126/sciadv.1700672}{{\em Sci. Adv.} {\bf 2017}, {\em 3},~e1700672}.
		
		\bibitem[110]{Knight89}
		Kim, M.S.; de~Oliveira, F.A.M.; Knight, P.L.
		\newblock Properties of squeezed number states and squeezed thermal states.
		\newblock \href{https://doi.org/10.1103/PhysRevA.40.2494}{{\em Phys. Rev. A} {\bf 1989}, {\em 40},~2494--2503}.
		
		\bibitem[111]{Puskarov16}
		Puskarov, T.; Schuricht, D.
		\newblock {Time evolution during and after finite-time quantum quenches in the 
		transverse-field Ising chain}.
		\newblock \href{https://doi.org/10.21468/SciPostPhys.1.1.003}{{\em SciPost Phys.} {\bf 2016}, {\em 1},~003}.
		
		\bibitem[112]{Cole13}
		Jeske, J.; Vogt, N.; Cole, J.H.
		\newblock Excitation and state transfer through spin chains in the presence of 
		spatially correlated noise.
		\newblock \href{https://doi.org/10.1103/PhysRevA.88.062333}{{\em Phys. Rev. A} {\bf 2013}, {\em 88},~062333}.
		
		\bibitem[113]{Takei18}
		Aftergood, J.; Takei, S.
		\newblock Noise in tunneling spin current across coupled quantum spin chains.
		\newblock \href{https://doi.org/10.1103/PhysRevB.97.014427}{{\em Phys. Rev. B} {\bf 2018}, {\em 97},~014427}.
		
		\bibitem[114]{Gangadharaiah21}
		Singh, M.; Gangadharaiah, S.
		\newblock Driven quantum spin chain in the presence of noise: Anti-Kibble-Zurek 
		behavior.
		\newblock \href{https://doi.org/10.1103/PhysRevB.104.064313}{{\em Phys. Rev. B} {\bf 2021}, {\em 104},~064313}.
		
		\bibitem[115]{Troyer05}
		Werner, P.; Troyer, M.; Sachdev, S.
		\newblock Quantum Spin Chains with Site Dissipation.
		\newblock
        \href{https://doi.org/10.1143/JPSJS.74S.67}{{\em J. Phys. Soc. Jpn.} {\bf 2005}, {\em 74},~67--70}.
		
		\bibitem[116]{HongGang15}
		Chen, C.; An, J.H.; Luo, H.G.; Sun, C.P.; Oh, C.H.
		\newblock Floquet control of quantum dissipation in spin chains.
		\newblock \href{https://doi.org/10.1103/PhysRevA.91.052122}{{\em Phys. Rev. A} {\bf 2015}, {\em 91},~052122}.
		
		\bibitem[117]{Giedke13}
		Schwager, H.; Cirac, J.I.; Giedke, G.
		\newblock Dissipative spin chains: Implementation with cold atoms and steady-state 
		properties.
		\newblock \href{https://doi.org/10.1103/PhysRevA.87.022110}{{\em Phys. Rev. A} {\bf 2013}, {\em 87},~022110}.
		
		\bibitem[118]{Katsura19}
		Shibata, N.; Katsura, H.
		\newblock Dissipative spin chain as a non-Hermitian Kitaev ladder.
		\newblock \href{https://doi.org/10.1103/PhysRevB.99.174303}{{\em Phys. Rev. B} {\bf 2019}, {\em 99},~174303}}.
\end{thebibliography}
\end{document}